\documentclass[preprint, 12pt]{aastex}

\begin{document}
\title{Host Galaxies of Hard X-ray Selected Type-2 Active Galactic Nuclei at
Intermediate Redshifts\altaffilmark{1}}
\author{Gaku Kiuchi\altaffilmark{2},
        Kouji Ohta\altaffilmark{2},
        Masayuki Akiyama\altaffilmark{3},
        Kentaro Aoki\altaffilmark{3},
        Yoshihiro Ueda\altaffilmark{2}
}
\email{gaku@kusastro.kyoto-u.ac.jp}
\altaffiltext{1}{Based on observations made with the University of Hawaii 
2.2 m telescope.}
\altaffiltext{2}{Department of Astronomy Kyoto University, Kyoto
606-8502, Japan}
\altaffiltext{3}{Subaru Telescope, National Astronomical Observatory of
Japan, 650 North A'ohoku Place, Hilo, HI 96720}
\begin{abstract}

We study properties of the host galaxies of 15 hard X-ray selected
type-2 active galactic nuclei (AGNs) at intermediate redshifts
(0.05$<z<$0.6) detected in $ASCA$ surveys. The absorption corrected
hard X-ray luminosities $L_{\rm 2-10 keV}$ range from 10$^{42}$ erg
s$^{-1}$ to $10^{45}$ erg s$^{-1}$.  We took the $R$-band image of
these AGNs with the University of Hawaii 2.2~m telescope. Thanks to
the intrinsic obscuration of nuclear light, we can decompose the
galaxies with a spheroid component and a disk component. The resulting spheroid
luminosities correlate with $L_{\rm 2-10 keV}$; higher (lower) X-ray
luminosity AGNs tend to reside in luminous (less luminous)
spheroids. It is also found that the hosts of luminous AGNs show a
large spheroid-to-disk luminosity ratio ($\sim$1), while those of less
luminous AGNs spread between 0 and 1. The correlation between $L_{\rm
2-10keV}$ and spheroid luminosity indicates that the relation between
mass of a supermassive black hole (SMBH) and spheroid luminosity
(BS-relation) at the intermediate redshifts. BS-relation agrees with that in the local universe, if the Eddington
ratio of 0.24 is adopted, which is a mean value determined from our $ASCA$
type-1 AGN sample at similar redshifts through the broad-line width
and continuum luminosity.   
The present study demonstrates the effectiveness of using type-2 AGNs at
high redshifts to study their host properties.

\end{abstract}
\section{Introduction}

It is now widely accepted that masses of supermassive black holes
(SMBHs) located at the centers of galaxies correlate with spheroid
(bulge) luminosities (or masses) of the host galaxies very tightly
(hereafter we call it ``BS-relation''; e.g., Kormendy and Richstone
1995; Magorrian et al. 1998; Ferrarese \& Merritt 2000; Gebhardt et
al. 2000; Marconi \& Hunt 2003). The tightness of the BS-relation
suggests an evolutional link between SMBHs and their spheroid
components of the hosts.  Marconi et al. (2004) and Shankar et
al. (2004) showed that a local black hole mass function derived from a
spheroid luminosity function by using the BS-relation agrees with that
expected from AGN relics estimated from the history of mass accretion
onto BHs during AGN phases. This suggests that all local spheroids
(bulges) have central BHs and have experienced an AGN phase (or
phases) during its life. Marconi et al. (2004) also suggested that the
cosmic mass accretion history onto BHs has a similar redshift
dependence to the cosmic star formation rate history, indicating that
the mass accretion rate onto a SMBH is proportional to the star
formation rate in terms of cosmological time scale.  These results
suggest the co-evolution of SMBHs and spheroids of host galaxies.

Using a highly complete hard X-ray selected AGN sample, Ueda et
al. (2003) revealed that the comoving number density of AGNs with high
X-ray luminosities shows a peak at higher redshift than that of low
X-ray luminosity ones. This indicates ``anti-hierarchical'' or
``down-sizing'' nature of AGN evolution. More recently, the same
tendency has been confirmed both in a study of soft X-ray selected
type-1 AGNs (Hasinger et al. 2005) and that of hard X-ray selected
AGNs (La Franca et al. 2005). Hard X-ray luminosities represent masses
of central SMBHs if the ratio of the X-ray luminosity to the Eddington
luminosity (Eddington ratio) is constant. From the co-evolution
scenario of SMBHs and spheroids, it is thus suggested that low X-ray
luminosity AGNs reside in late-type galaxies having a small spheroid
component, while high X-ray luminosity AGNs reside in early-type
galaxies having a luminous spheroid. If this is the case, the number
density evolution of AGNs indicates that more massive spheroid
galaxies and BHs both formed at earlier epoch.

A direct approach to examine the co-evolution of SMBHs and spheroids
is to determine BS-relation at $z>0$ and compare it with the local
relation. Host galaxies of type-1 AGNs have been studied extensively
(e.g., Sanchez et al. 2004; Jahnke et al. 2004). It is not easy,
however, to examine morphologies and luminosities of host galaxies,
particularly of their spheroid (bulge) components, because the
presence of a dazzling nucleus prevents us from examining a host
galaxy precisely. Meanwhile, type-2 AGNs make a suitable sample to
study structure of host galaxies thanks to its intrinsic obscuration
of a bright nucleus. According to the AGN unification model, type-1
and type-2 AGNs are intrinsically the same population; the only
difference is the viewing angle to a nucleus surrounded by dusty
obscuring material. Hence, examining the properties of the host
galaxies of type-2 AGNs should be useful for understanding those of
the whole AGN population.

However, since previous AGN surveys were mainly conducted in UV/optical or soft
X-ray bands, detection of obscured AGNs (type-2 AGNs) has
been difficult due to extinction and absorption. Hard X-ray surveys
have much less bias against obscured AGNs. Akiyama et al. (2000) and
Akiyama et al. (2003) completed optical follow-up observations of hard
X-ray sources detected with $ASCA$ at a flux limit of $\sim 10^{-13}$
erg cm$^{-2}$ s$^{-1}$ (2--10 keV), providing us with a unique type-2
AGN sample at intermediate redshifts ($0.05<z<0.6$). In this paper, we
use this $ASCA$ type-2 AGN sample to examine the relation between
X-ray luminosity, spheroid luminosity, and bulge-to-total luminosity
ratio (B/T ratio).

The sequence of the article is as follows. In Section~2 we describe
the sample used in the analysis. The observation and data reduction
are summarized in Section 3. In Section~4, we explain the procedure of
two-dimensional surface brightness fitting. The results are presented
in Section~5, and summary is given in Section~6.
Throughout this paper, we adopt a cosmological parameter set of 
$H_0=70$ km s$^{-1}$ Mpc$^{-1}$, $\Omega_{m}=0.3$, and
$\Omega_{\Lambda}=0.7$, and the Vega system for magnitudes unless
otherwise noted. 

\section{Sample}

We utilize two hard X-ray ($>2$ keV) selected samples, the $ASCA$
Large Sky Survey (ALSS) and $ASCA$ Medium Sensitivity Survey in the
northern sky (AMSSn). The ALSS covers a contiguous area of $\approx$5
deg$^{2}$ near the north Galactic pole with the flux limit of about
$1\times$ 10$^{-13}$ erg s$^{-1}$ cm$^{-2}$ in the 2--10 keV band
(Ueda et al.\ 1999). Thirty-four sources were detected above 2 keV
with the Solid-state Imaging Spectrometer, among which 33 sources were
optically identified into 30 AGNs, 2 clusters of galaxies, and 1
Galactic star (Akiyama et al.\ 2000). The AMSSn is a serendipitous
source survey with the Gas Imaging Spectrometer (Ueda et
al. 2001). From a survey area of $\sim$ 70 deg$^{2}$ with a flux limit
of $3\times$ 10$^{-13}$ erg s$^{-1}$ cm$^{-2}$ (2--10 keV), all but
one of 87 hard X-ray selected sources were identified with 78 AGNs, 7
clusters, and 1 Galactic object (Akiyama et al. 2003). Because these
AGNs are selected in the hard X-ray band, they are not biased to
unabsorbed AGNs, except for Compton thick AGNs. It is worth
emphasizing that the completeness of the sample is almost 100\%, and
hence we can make a statistically fair sample of AGNs at intermediate
redshifts.

\begin{figure}
 \plotone{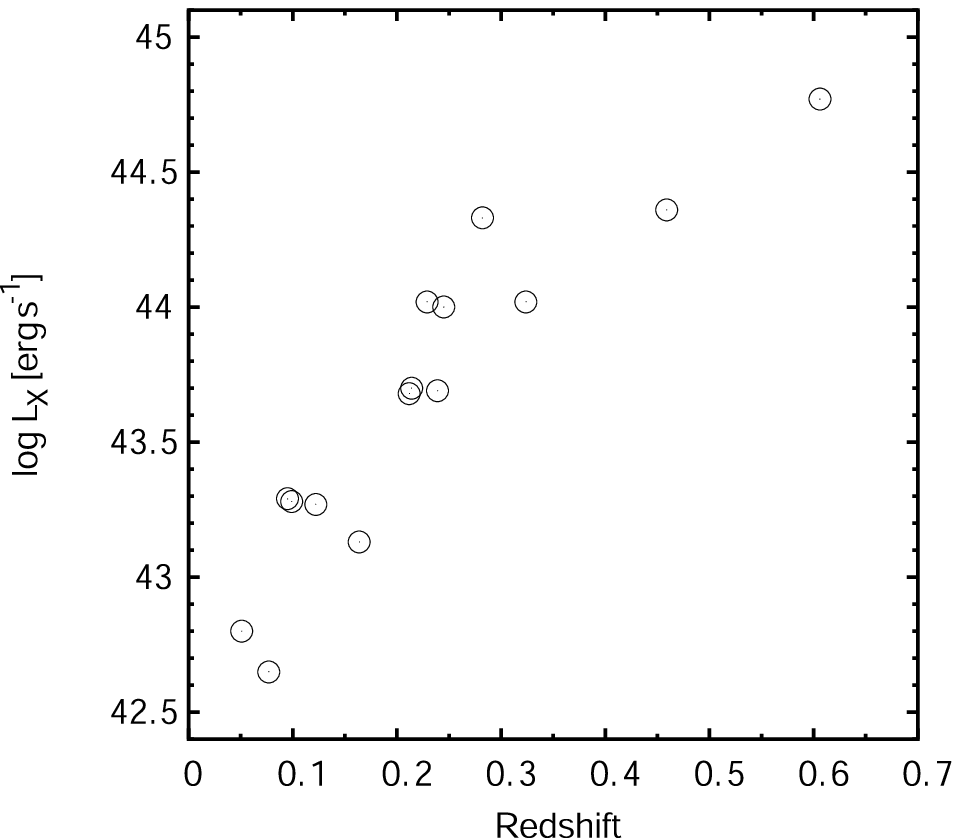}
 \caption{Absorption corrected X-ray luminosities in the rest frame 2--10
 keV band against redshifts of our sample.\label{z_lx}}
\end{figure}

For our study, we select only type-2 AGNs from the identified $ASCA$
sources. The definition of ``type-2'' AGNs we adopt is the ordinary
criteria; neither H$\alpha$ nor H$\beta$ broad emission lines are seen
in the optical spectrum. Two objects (NE04 and NO53) are included to
our sample based only on the absence of the H$\beta$ broad emission
lines, because their H$\alpha$ emission line is not covered in the
optical spectra taken. Although the absence of broad line(s) could
depend on the signal-to-noise ratio, we confirm that these AGNs show
absorption of $N_{\rm H} \gtrsim 10^{22}$ cm$^{-2}$ in the X-ray band,
thus supporting their identification as type-2 AGNs. We do not include
X-ray absorbed ``type-1'' AGNs (i.e., with optical broad emission
lines), which tend to have a bright nucleus in the optical band.  The
number of the total type-2 AGNs obtained from the ALSS and AMSSn is
15. The basic properties of the sample are summarized in Table
\ref{sample}. The $R$-band magnitude ranges between 14.8--20.6 mag, and the
absorption-corrected X-ray luminosity ranges from 10$^{42.7}$ erg
s$^{-1}$ to 10$^{44.8}$ erg s$^{-1}$ (2--10 keV). Their redshifts
distribute from 0.05 to 0.6 with a median redshift of 0.22. Figure
\ref{z_lx} shows the plot of X-ray luminosity against redshift for
this sample.

\begin{figure}
 \plottwo{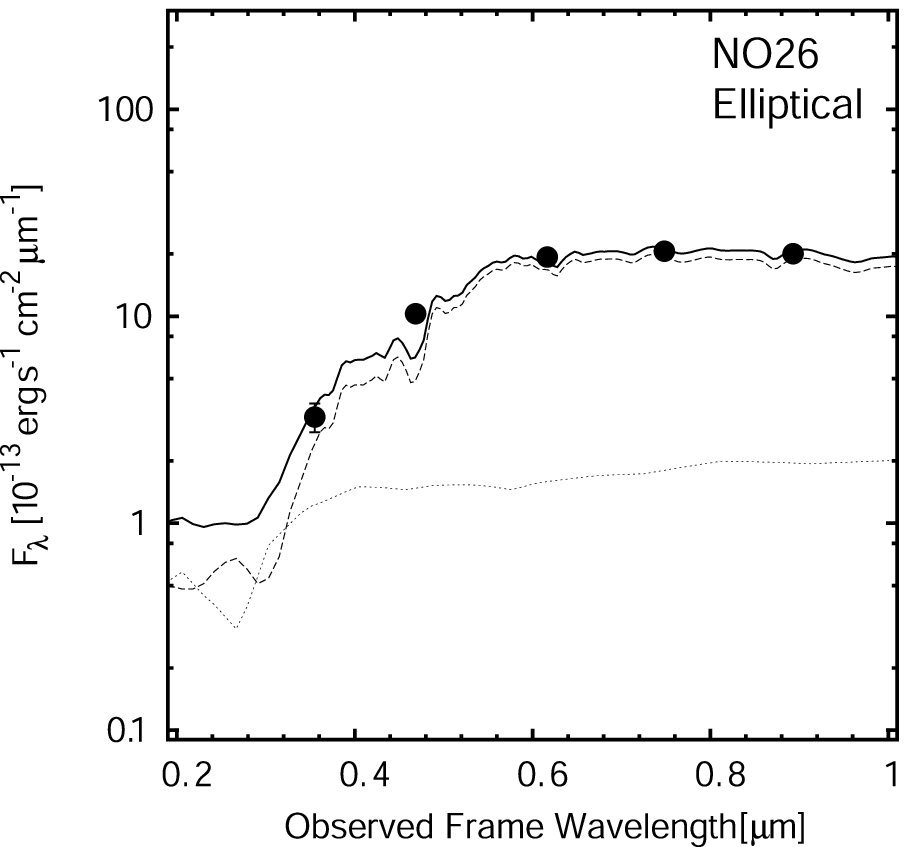}{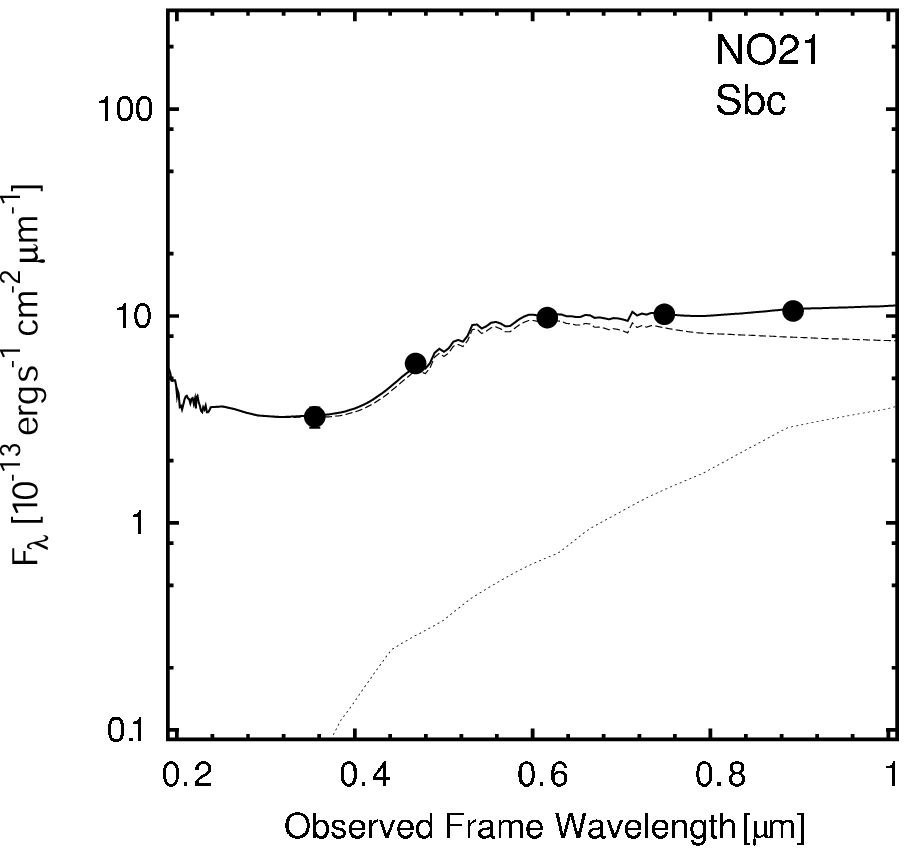}
 \caption{Spectral energy distributions (SEDs) of NO26(left) and
 NO21(right) taken from the SDSS data (filled circles). Dashed, dotted, and solid
 lines show a spectrum of a galaxy, a nuclear component, and a total,
 respectively. The nuclear component does not dominate observed optical
 light, and its contribution is less than $\sim$ 10 \% in $r$-band. \label{sed}}
\end{figure}

The possible contribution from a nuclear point source (via direct
light and/or scattered light from the nucleus) is negligible in the
optical band in our sample. A part of the sources show
Mg$_{b}\lambda$5175, Ca H , and Ca K absorption lines in the spectra,
indicating that the optical luminosity is likely to be dominated by
stars in the host galaxy. Furthermore, 10 objects in this sample have 
photometric data from the Sloan Digital Sky Survey (SDSS),
by which we can examine their spectral energy distribution
(SEDs). Figure \ref{sed} shows typical examples of the resultant
SEDs. These SEDs are similar to that of elliptical or Sbc galaxies,
suggesting that the contribution from a nuclear component is small. To
evaluate this quantitatively, we fit the SEDs with a combined spectral
model of a galaxy and an AGN. Here we use three template spectra of
Elliptical, Sbc, and Scd (Coleman, Wu, \& Weedman 1980) to reproduce
the host galaxy spectra, and a mean QSO spectrum by Elvis et
al. (1994) to reproduce the nucleus spectrum with A$_{V}$ = 0--30 mag
by employing the extinction curve of Milky Way (Allen 1996). The
fitting results are also shown in Figure \ref{sed}; the contribution
from the nucleus is less than 10\% at $r$-band. Small nuclear
contribution ($<10$\%) is also confirmed for the other objects that
have SDSS photometric data.

\section{Observations and Data Reductions}

Optical imaging observations were made during the period from 2004
April 16 to April 19 and on December 10 with the Orthogonal Parallel
Transfer Imaging Camera (OPTIC, Tonry et al. 2004) attached to the
University of Hawaii 2.2 m (UH88) telescope. The OPTIC can compensate
for real time image motion by moving collected charges in the CCD
array in response to on-chip guide stars. One pixel corresponded to
0$^{\prime\prime}$.14, giving a field of view of $\sim 5\times5$
arcmin$^{2}$. In the April observing run, about 20 targets were
observed. The integration time of each frame was 600s--900s in
$R$-band and the total exposure time was 600s--1800s. The typical
seeing size (FWHM) was 1$^{\prime\prime}$.5, and the sky condition was
not photometric. In the December observing run, we observed one
target. The integration time of each frame was 600s in $R$-band and
the total exposure time was 1800s. The typical seeing size was
0$^{\prime\prime}$.8. The sky condition was photometric, and Landolt
standard stars were also observed for the photometric calibration.
Since the observing conditions for the spring run were bad, we again
made observations on 2005 May 12 and 13 at UH88 with Tek2048 CCD to
obtain images of most of the spring targets. One pixel of this CCD
corresponded to 0$^{\prime\prime}$.22, giving a field of view of
 $\sim$ 7.5$\times$7.5 arcmin$^{2}$. The integration time of each frame was
600s$-$900s in $R$-band, and the total exposure time was 600s--1800s.
The typical seeing size was 0$^{\prime\prime}$.8 during the
observations. Since the sky condition was photometric on the first
night, we observed standard stars on that day. A journal of the
observations is given in Table~\ref{sample}.

\begin{figure}
 \plotone{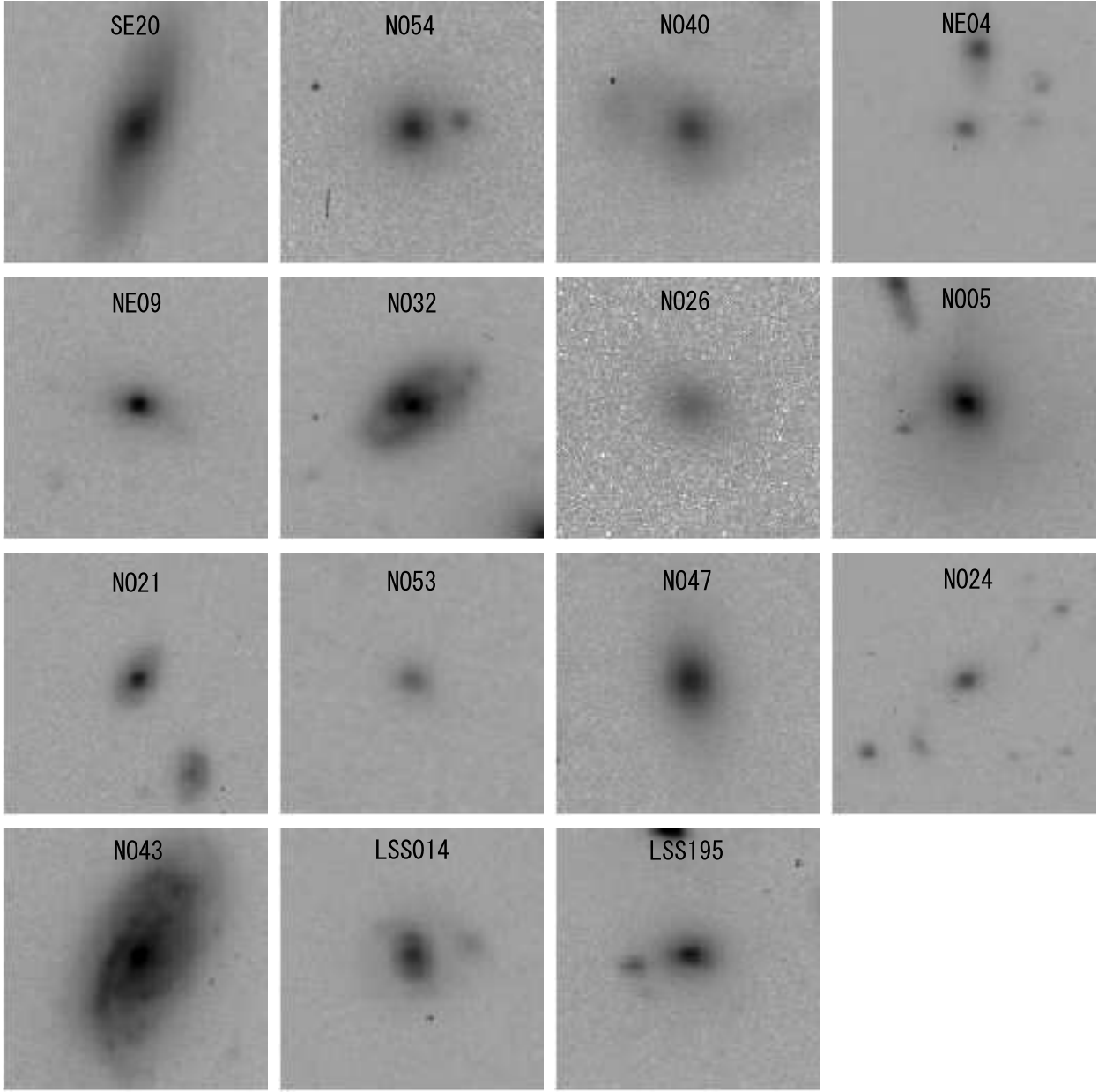}
 \caption{$R$-band images of the targets. Each image covers 25$^{\prime\prime}$
 $\times$ 25$^{\prime\prime}$. North is at the top and east is to the left.\label{img}}
\end{figure}

We reduced the data taken with the OPTIC by standard procedure except
for flat fielding, for which we created custom flat fields by
convolving the average dome flat frames with a shift pattern of
charges in each exposure. For the reduction of the data taken with Tek
2048, we took standard procedure. We subtracted bias values estimated
with a median of the overscan region. Bias-subtracted object frames
were divided by flat frame, which was taken from a median frame of
normalized dome flat frames. A sky background was determined as the
mode value of 300 pixels $\times$ 300 pixels around a target, and was
subtracted from the image. By using the positions of bright stars in a
frame, 1--3 frames for each target with a small offset were scaled and
combined into one frame by taking a median value. The resulting images
of the targets are shown in Figure \ref{img}.

Photometric calibration was made for the data taken in 2004 December
and 2005 May. The photometry of standard stars was performed with the
``PHOT'' task in the IRAF package\footnote{IRAF is distributed by the
National Optical Astronomy Observatory, which is operated by the
Association of Universities for Research in Astronomy, Inc. (AURA),
under cooperative agreement with the National Science Foundation.}.
After examining the growth curve of the targets, we set an aperture
size to three times the FWHM of each object, which is 
appropriate to derive the total magnitude by avoiding contamination from nearby
objects. The uncertainty of the obtained zero point of magnitude is
$\pm$ 0.03 mag. For the objects for which SDSS data are available, we
calculate the $R$-band magnitudes from the SDSS $r$-band magnitudes by
using the transformation in the SDSS web
page\footnote{http://www.sdss.org/dr4/algorithms/sdssUBVRITransform.html\#Lupton2005},
 i.e.,
\begin{equation}
R [{\rm Vega}] = r[{\rm AB}] - 0.1837(g[{\rm AB}] - r[{\rm AB}]) - 0.0971 - 0.212,
\end{equation}
where the last term 0.212 corresponds to the AB magnitude of $\alpha$
Lyrae at the band.
We confirm that the $R$-band magnitudes converted from the SDSS data
agree with our photometric results within $\sim$ 0.2 mag. 
For the remainder objects, we adopt $R$-band magnitudes 
given by Akiyama et al. (2003) and Watanabe et al. (2004); 
these values were derived by converting $O$ and $E$ magnitudes
in the APM catalog into $B$ and $R$ magnitudes.
The $R$-band magnitudes obtained in this way are summarized in Table
\ref{sample}. Finally, we derived K-corrected absolute $R$-band
magnitudes using model spectra fitted to SEDs for the objects for
which SDSS data are available, and the Sbc template spectrum for the
remainders. The resultant absolute magnitudes are also listed in Table
\ref{sample}.

\section{Surface Brightness Fitting}

We performed two-dimensional fitting to the $R$-band images to
determine spheroid (bulge) and disk luminosities by using the
``GALFIT'' program (ver 2.0.3, Peng et al. 2002). The model consists
of three components, a de Vaucouleurs $r^{1/4}$ spheroid (bulge), an
exponential disk, and a nuclear point-like component which may be
present even in type-2 AGNs. A point spread function (PSF) was
constructed by stacking stellar
images of 1--3 bright (but not saturated) stars in the same image that
contains the AGN, and is used as the nuclear component.  The model of
the spheroid and disk components are convolved with this PSF to
incorporate the seeing effect. The fitting parameters are a position
that is common for the three components, magnitudes of each component,
scale lengths, ellipticities, and position angles for the spheroid and
disk components.

%least square method = \sigma (Data-model)^2/error^2 の説明はいりますかね？
%いらないような気はしますが。

To search for the best-fit parameters, we employed the least square
method. Considering possible systematic errors, we set a constant
fractional error in each pixel regardless of the photon counts. Objects
close to the target, if any, were masked out. To converge the fit
efficiently, we took two steps. First, we derived a position, a total
magnitude, an ellipticity, and a position angle of a target galaxy, by
using the IMEXAM task in IRAF, to be used as initial parameters. We
set 1 kpc and 3 kpc for the scale length of a spheroid and a disk,
respectively, and a B/T ratio of 0.5. With this initial guess, fitting
was performed without a nuclear component, since the nuclear
luminosity can be regarded to be small compared with the host galaxy
luminosity as zeroth-order approximation. In the second step, we
iterate the fitting by including a nuclear component, whose flux was
initially set to be 10\% of the total flux.

\begin{figure}
 \plotone{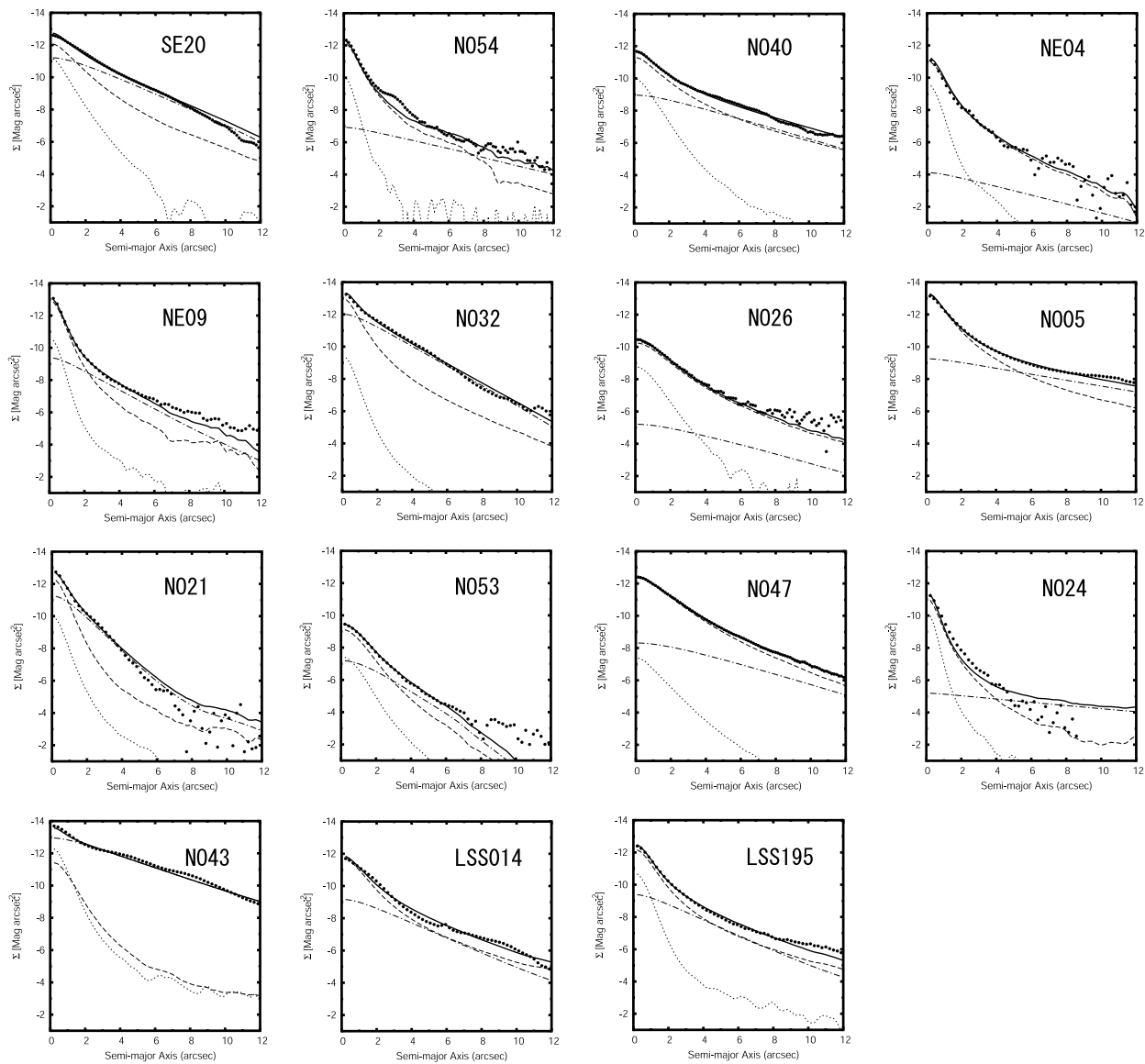}
 \caption{Azimuthally averaged radial profiles of surface
 brightness. Vertical axis is arbitrary. Dots are data points. 
Dotted line, dashed  line, dotted-dash line, and solid line show 
PSF, de Vaucouleurs $r^{1/4}$ spheroid, exponential disk, 
and total components of the best fit model, respectively.
% Surface brightness is plotted with relative scale.
 \label{prof}}
\end{figure}

\begin{figure}
 \plotone{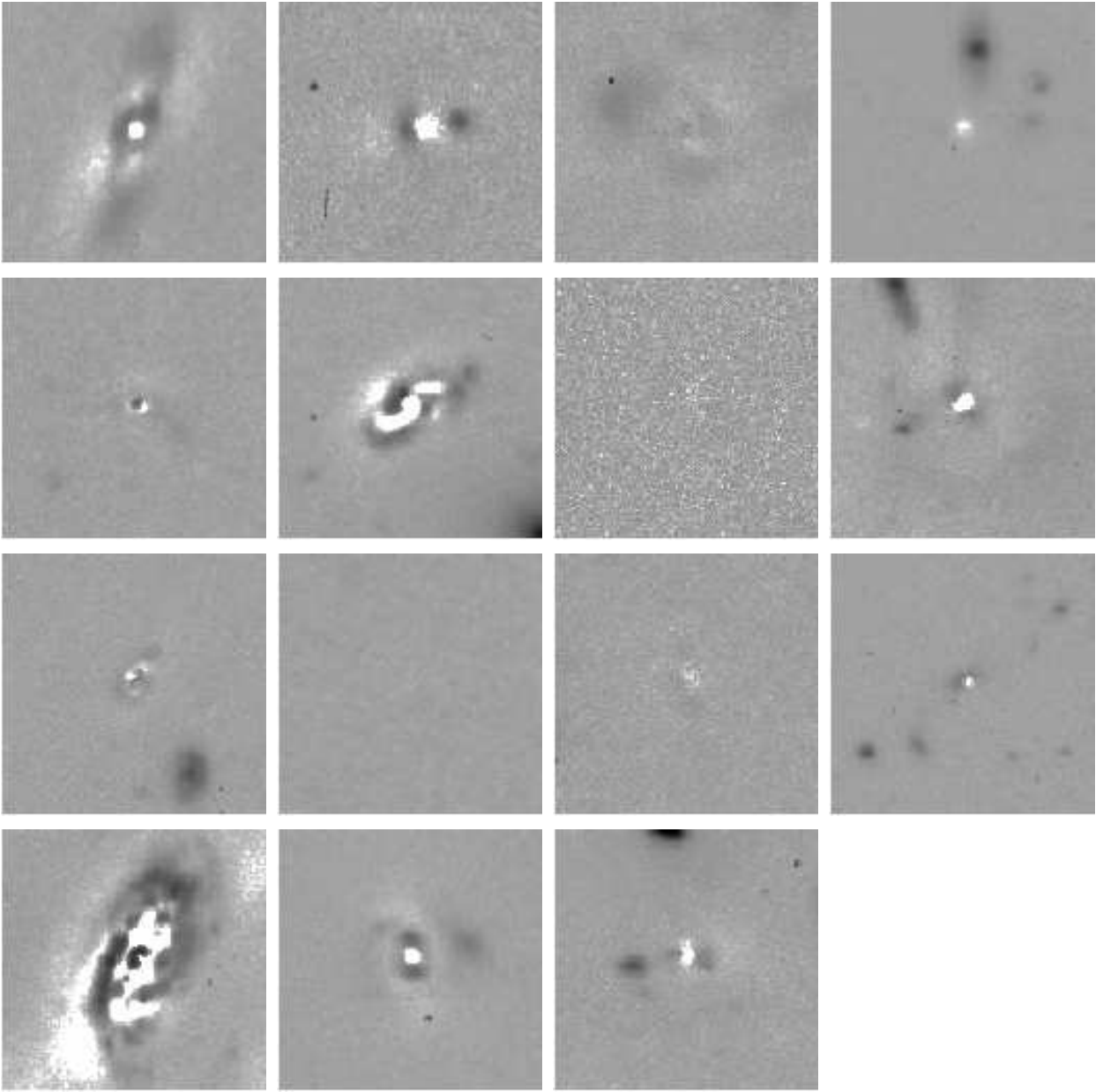}
 \caption{$R$-band residual images. Each image covers 25$^{\prime\prime}$
 $\times$ 25$^{\prime\prime}$. Images are displayed in the same order as Figure
 \ref{img}. Display range is the same as Figure \ref{img}.\label{res}}
\end{figure}

The obtained best-fit parameters are summarized in
Table~\ref{fit_lst}. Figure~\ref{prof} shows the radial
surface-brightness profile together with the best-fit model. 
We note that NO26 is not clearly decomposed to a spheroid and a disk
 component because of the large seeing size.
 The residual images of the fitting are shown in
 Figure~\ref{res}. Asymmetric structures, such as
a bar or an arm, are seen in some objects.  Some objects show a hint
for the presence of companions.  Since no redshift data are available
for them, however, the physical relation is not certain.

We performed simulation to confirm the robustness of our results and
to evaluate the uncertainties in the best-fit parameters. For each
object, we produced simulated images of the galaxy consisting of the
three components, a spheroid, a disk, and a nucleus, assuming the
best-fit parameters. 
% The images were convolved with the point spread
%function with the same seeing condition of the actual observation.
 The images were convolved with the same point spread function as the 
actual observations. The background was taken from a blank sky region 
in the observed image. By adding Poisson noise in the photon counts, we
produced 50 simulated
images per object with different random numbers. Also, to evaluate
possible systematic errors caused by the complexity in the background,
we repeated the simulation by selecting other blank sky regions in the
observed image. Then, we performed image fitting to the simulated
images according to the same procedure applied to the actual data, and
compared the output parameters with the input values. The distribution
of the output parameters give us an estimate of the uncertainty in the
parameters including both statistical and systematic errors.

Based on this study, we conservatively estimate the uncertainties in
the spheroid magnitude to be 0.2 mag for most of the objects. The
objects NO43 and NO26 show larger errors, 0.5 mag and 0.3 mag,
respectively. This is because NO43 has a very weak spheroid component
compared with the disk component, making the coupling from the nuclear
component quite significant. As mentioned above, the seeing condition
of NO26 was extremely poor (1$^{\prime\prime}$.5), which causes similar
difficulty in separating different galactic components. The
uncertainties of the B/T ratio are found to be 0.1 for NO26, and 0.05
for the rest (NO43 is a disk dominated object, hence error of its B/T ratio
is small). We take into account these uncertainties in the
following discussions.

\section{Relation between the AGN and Host Properties}
\subsection{Correlation between X-ray luminosity and spheroid luminosity}
%correlation

\begin{figure}
 \plotone{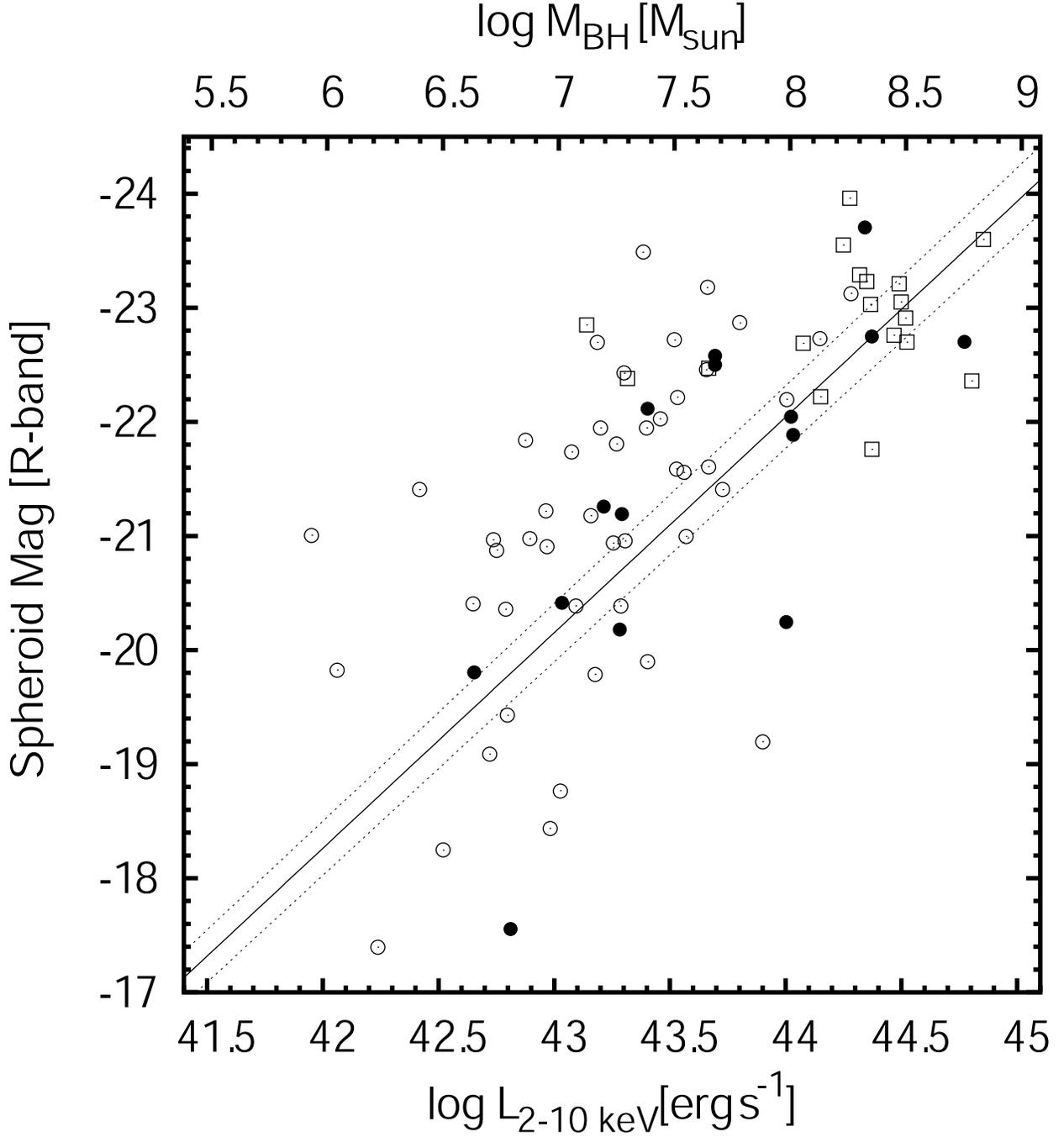}
 \caption{Absolute spheroid magnitude in $R$-band against $L_{\rm 2-10keV}$(filled circles).
 A typical error of $L_{\rm 2-10keV}$ and the spheroid magnitude
 are estimated to be 0.1 dex and 0.2--0.5 mag, respectively. Upper ordinate indicates masses of
 SMBH estimated from $L_{\rm 2-10keV}$ by assuming $BC=30$(bolometric
 correction) and
 $\lambda=0.24$(Eddington ratio). Solid and dotted lines represent the best
 fit model and a 1$\sigma$ uncertainty, respectively.  
 Open circles and open squares refer to data by Schade et al. (2000) and
 by Dunlop et al. (2003).  
Values of $L_{\rm 2-10keV}$ for the samples of Schade et al. (2000) and 
Dunlop et  al. (2003)  are derived as described in the text.
\label{bs_agn2}}
\end{figure}

In Figure~\ref{bs_agn2} we plot the absolute magnitudes of spheroid
components against absorption-corrected X-ray luminosities in the
rest-frame 2--10 keV band. As seen from the figure, the absolute
magnitudes correlate with the hard X-ray luminosities; i.e., high
(low) X-ray luminosity AGNs reside in more (less) luminous spheroid
components. If we assume that the Eddington ratio is constant,
this correlation indicates the presence of a BS-relation at
$z=0.05-0.6$ similar to that in the local universe.
The mass of a SMBH, $M_{\rm BH}$, can be derived from the X-ray
luminosity $L_{\rm 2-10 keV}$ as
\begin{equation}
\log \left(\frac{M_{BH}}{M_{\odot}}\right) = \log \left(\frac{L_{\rm 2-10keV}}
{\rm erg\ s^{-1}}\right) + \log \left( \frac{BC}{30} \right) - 
\log \left( \frac{\lambda}{0.1}  \right) - 35.6,
\end{equation}
where $\lambda$ is the Eddington ratio and $BC$ is a bolometric
correction factor, for which we adopt 30 from a mean spectrum of
quasars compiled by Elvis et al.\ (1994).

\begin{figure}
 \plotone{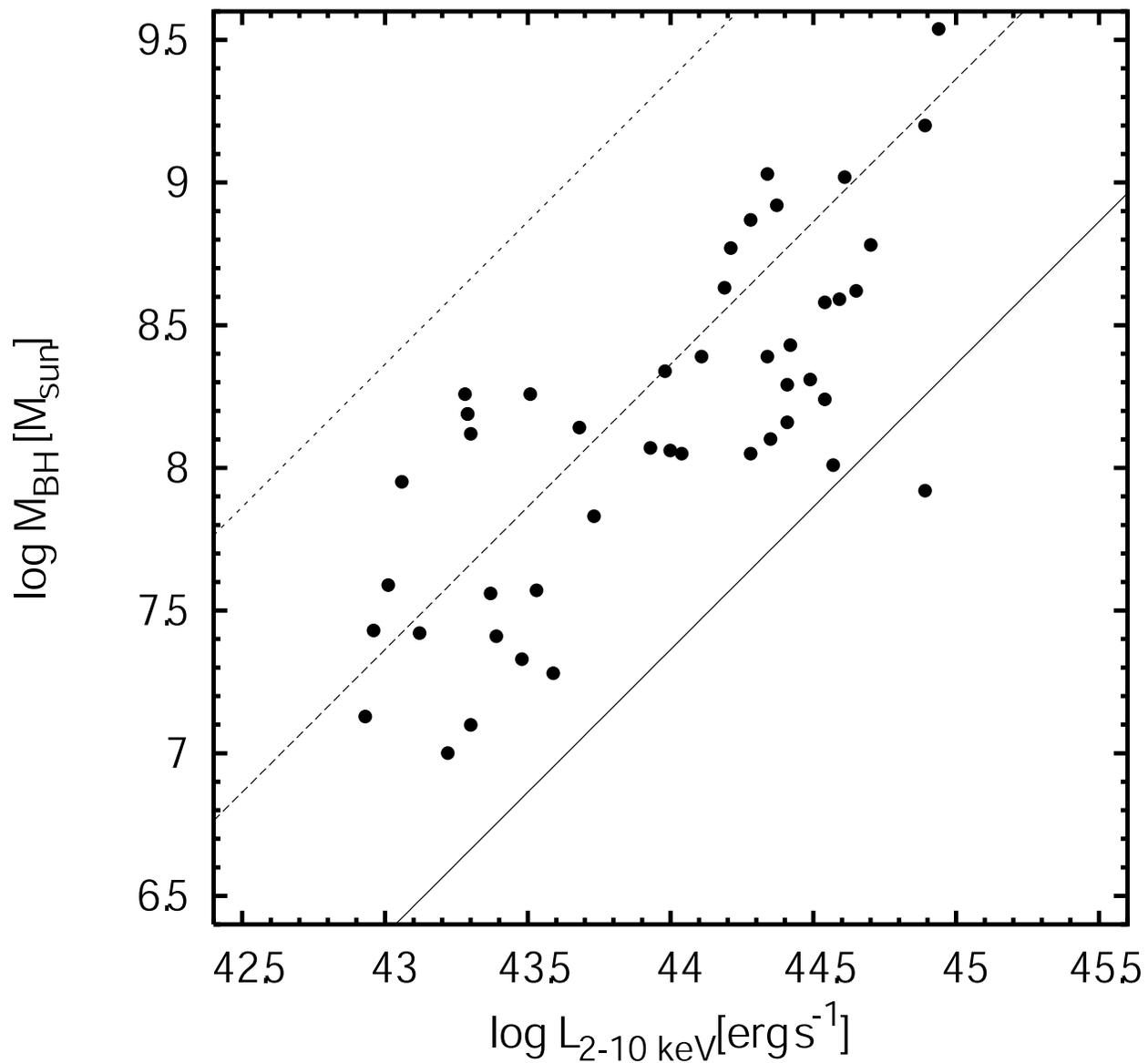}
 \caption{$M_{BH}$ against $L_{\rm 2-10keV}$ for type-1 AGNs in ALSS and AMSSn.  
 $M_{BH}$ is derived by Kaspi relation. Solid, dashed, and dotted
 lines represent the Eddington ratio of 1, 0.1, and 0.01,
 respectively.\label{mbh12}}
\end{figure}

To estimate $\lambda$, we utilized type-1 AGNs from the ALSS and AMSSn
sample. Their redshifts and hard X-ray luminosities range from 0.1 and
0.7, and from 10$^{43}$ to 10$^{45}$ erg s$^{-1}$, respectively, which
are almost the same as those of our type-2 AGNs. According to the
unified scheme, we can expect that the Eddington ratio should be same
between the type-1 and type-2 AGNs in average. Figure~\ref{mbh12}
shows the correlation between the $L_{\rm 2-10 keV}$ and black hole
mass of the type-1 AGNs. The masses are derived through the Kaspi
relation (Kaspi et al. 2000) from the velocity width of H$\beta$ broad
line and continuum luminosity summarized in Akiyama et al.\ (2000,
2003). Overall, there is a good correlation between the two, with a
scatter of $\approx$0.5 dex around the mean value of $\lambda = 0.24$,
being consistent with other studies (e.g., Kollmeier et al. 2005).
This result supports our assumption that the Eddington ratio can be
regarded to be constant in our sample at first-order approximation.

Now we derive a BS-relation of the type-2 AGNs by converting 
$L_{\rm 2-10 keV}$ into $M_{\rm BH}$ assuming $\lambda=0.24$. 
The upper ordinate of Figure~\ref{bs_agn2} shows the mass of SMBHs thus obtained. 
The solid line represents the best-fit BS-relation, 
\begin{equation}
M_{R\ {\rm sph}} = -1.8(\pm 0.1)\log\left( \frac{M_{BH}}{M_{\odot}}\right)-6.9(\pm0.7),
\end{equation}
where $M_{R\ {\rm sph}}$ is the $R$-band spheroid absolute magnitude. 
McLure \& Dunlop (2002) also show the presence of a BS-relation 
for a sample of inactive galaxies, type-1 quasars, and Seyfert galaxies at $z<0.5$
by using HST imaging data, described as $M_{R\ {\rm sph}} = -2.0(\pm
0.1)\log\left(\frac{M_{BH}}{M_{\odot}}\right)-5.92(\pm0.99)$, 
with a cosmological parameter set of $H_0=50$ km s$^{-1}$ Mpc$^{-1}$ and $\Omega_{m}=1$.
Our BS-relation (equation 3) is converted to 
 $M_{R\ {\rm sph}} = -1.8(\pm
 0.1)\log\left(\frac{M_{BH}}{M_{\odot}}\right)-7.4(\pm0.8)$   
with the cosmological 
parameter set they adopted, hence both relations agree with each other
within the uncertainty.
Studying host galaxies of type-1 AGNs at such redshift range is
not easy, and generally HST images are needed to derive the spheroid
luminosities. Our study demonstrates that studying AGN hosts at
intermediate redshifts is possible even with a ground-based telescope,
if we utilize a type-2 AGN sample.

To compare our results of $L_{\rm 2-10 keV}$ - $M_{R\ {\rm sph}}$
relation with other studies made for type-1 AGNs, we also plot the
samples of Schade et al. (2000) and Dunlop et al. (2003) in Figure
\ref{bs_agn2} by open circles and open squares, respectively. Schade
et al. (2000) studied the host galaxies of soft X-ray selected type-1
AGNs at $z\sim 0.1$ by using the HST images. We converted the
monochromatic luminosity at 2 keV given in Schade et al. (2000) into
$L_{\rm 2-10 keV}$ by assuming an X-ray spectrum of $F_{\nu} \propto
\nu^{-0.9}$. Dunlop et al. (2003) also studied the host galaxies of
type-1 AGNs at $z < 0.3$ with the HST data. The conversion from the
$R$-band nuclear luminosity into the $L_{\rm 2-10 keV}$ is made with
the relations $\alpha_{\rm OX} = 0.1152 L_{2500}[{\rm erg\ s^{-1}\
Hz^{-1}}] - 2.0437$ (Ueda et al. 2003), $F_{\nu} \propto \nu^{-0.44}$
in the optical band, and $F_{\nu} \propto \nu^{-0.9}$ in the X-ray
band.  As seen in Figure \ref{bs_agn2}, the distribution of our sample
agrees well with that of the Dunlop et al. (2003) sample. However, it
is significantly different from that by Schade et al.\ (2000); the
spheroid luminosities against $L_{\rm 2-10 keV}$ in the Schade et al.\
(2000) sample are systematically higher, and have a larger scatter than
in our sample, particularly in the low X-ray luminosity range.  The
cause for the different distribution is not clear. Since it is
suggested that low luminosity AGNs tend to have lower Eddington ratios
with a larger scatter than high luminosity QSOs (e.g., Kollmeier et
al. 2005), it might be possible that the sample of Schade et
al. (2000) includes objects with lower Eddington ratios, while our
sample does not contain such objects.

\subsection{The correlation between B/T and nuclear luminosities}

\begin{figure}
 \plotone{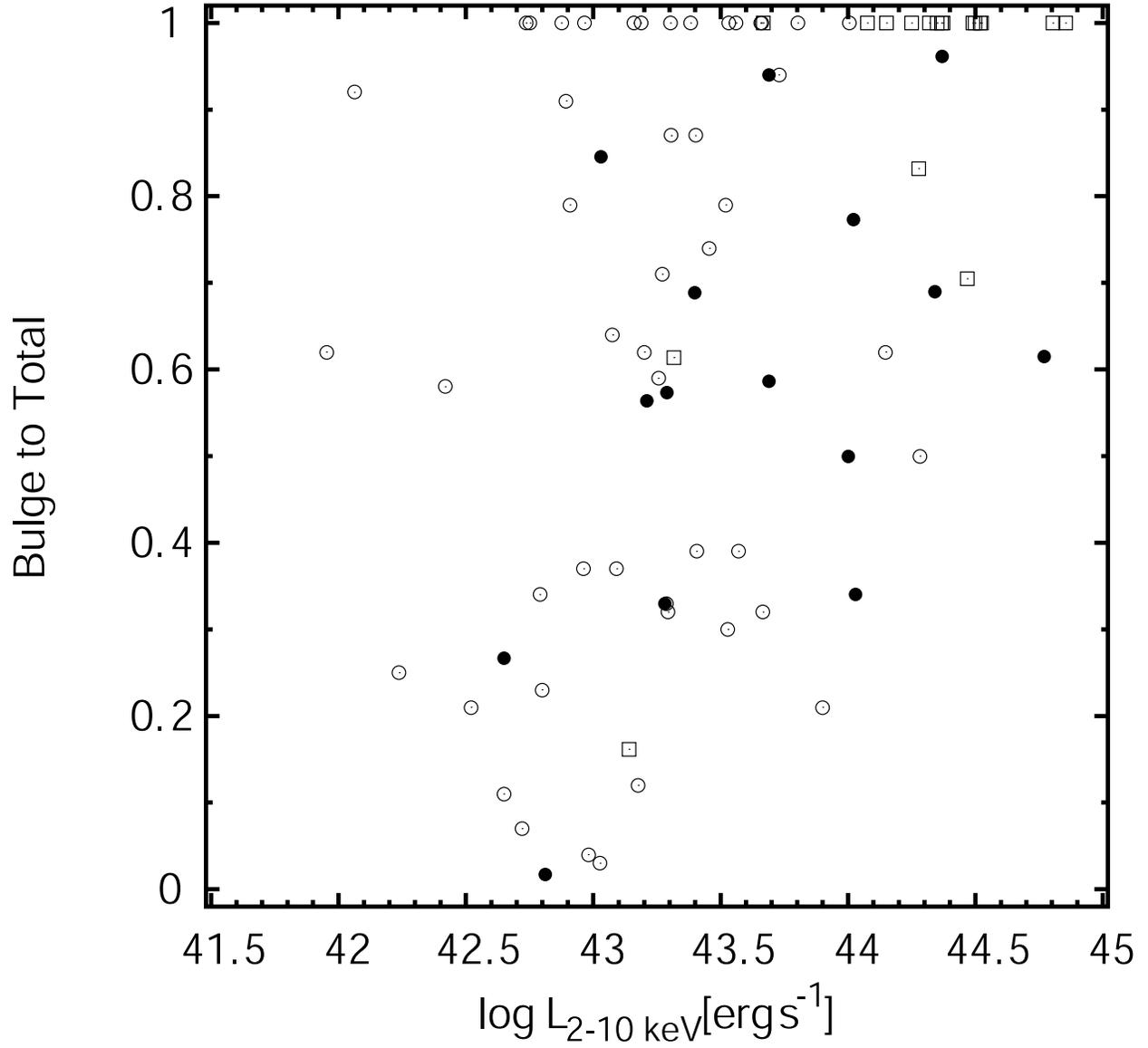}
 \caption{Spheroid to total luminosity ratio (B/T ratio) against
 $L_{\rm 2-10keV}$. A typical error of the B/T ratio is $\lesssim$
 0.1. Symbols are the same as Figure \ref{bs_agn2}. \label{bt}}
\end{figure}

The spheroid-to-total luminosity ratio (B/T ratio) is a key parameter
representing the morphology type of host galaxies. We plot the B/T
ratio against $L_{\rm 2-10 keV}$ for our sample by filled circles in
Figure~\ref{bt}, where the total luminosity refers to a sum of the
best-fit model spheroid luminosity and disk luminosity.
 As noticed from the figure, high X-ray luminosity
AGNs resides only in galaxies with the large B/T ratio, while low
X-ray luminosity AGNs reside in those with the B/T ratio of 0--1.0.
 If we take the total luminosity as the total integrated magnitude
 instead  of taking the sum of the best-fit model luminosities,
the values of B/T change about less than 0.05 (and up to to 0.1 for a few
objects). Thus the trend does not change. 
 We also plot the results of Schade et al. (2000) and Dunlop et
al. (2003) in Figure \ref{bt} by open circles and open squares,
respectively. The distribution of both samples is similar to that of
our sample; no AGNs reside in a region of high X-ray luminosity {\it
and} low B/T ratio. Some objects of Schade et al. (2000) and Dunlop et
al. (2003) show the B/T ratio of unity. This is because they added a
disk component to a model only when an observed image could not be
well reproduced with a spheroid plus a nuclear component. Thus, if the
same fitting method as used in this paper were applied to their data,
the B/T ratio would become slightly lower than unity. 
 Similarly, some objects with a high B/T ratio of our sample (NO54,
NE04, and NE09) could be reproduced without a disk component. 
Their spheroid luminosities become slightly higher, but these are within
uncertainties of the surface brightness fitting. 
In fact, B/T ratios do not change so much and the correlation
between B/T ratio and $L_{\rm 2-10 keV}$ still holds. 

\section{Summary}
We study properties of the host galaxies of 15 hard X-ray
selected type-2 AGNs at intermediate redshifts
$(z=0.05-0.6$ with a median redshift of 0.22), obtained from optical
identification (ALSS, AMSSn) of the sources detected in $ASCA$ surveys. 
Spheroid (bulge) and disk
luminosities of the host galaxies were derived by two-dimensional
surface brightness fitting.
We found the correlation between $L_{\rm 2-10keV}$ and spheroid
luminosity, implying the presence of the BS-relation at
the intermediate redshifts.
We estimated the Eddington ratios from type-1 AGNs
in our original sample of ALSS and AMSSn, and found that
the ratios distribute from 0.1 to 1.0 with the mean value of $\sim$ 0.24. 
Adopting the mean Eddington ratio of 0.24, we derived
a BS-relation at the intermediate redshifts, which agrees
with the BS-relation at the similar redshifts
 obtained by McLure and Dunlop (2002). 
The BS-relation obtained by McLure and Dunlop (2002)
and thus that obtained here show almost no evolution
from the BS-relation obtained at $z\sim0$.
Tamura et al. (2006) compared the BS-relation by McLure and Dunlop
(2002) with those for local galaxies by converting the photometric band, and showed that
the BS-relation almost agrees 
with local relations in $B$-band (Marconi and Hunt 2003;
Ferrarese and Merritt 2000) and in $V$-band (Merritt and Ferrarese
2001). 
We also examined the distribution of B/T ratio against $L_{\rm 2-10keV}$,
and found that B/T ratios of the X-ray luminous AGNs have large B/T ratios
close to 1.0, and those of less luminous AGNs have B/T ratios of 0 to 1.

%However, the BS-relation by McLure and Dunlop (2003) seems to
%show a slight shift ($\sim 0.2$ dex in log $L_B$)
% relative to the local relations. 
%Since the band conversion does not include the effect of
%passive evolution, this may be caused by passive evolution(or may be due
%to K-correction).
%The expected shift in luminosity due to passive evolution is
%$\sim 0.4$ mag in $B$-band from $z=0.3$ to 0 (Tamura et al. 2006),
%thus the slight shift might indicate the effect of the passive evolution.

%In addition, if $L_{\rm 2-10keV}$s could indicate B/T ratios, more massive BHs and more luminous
%early-type host galaxies are evolved at earlier epoch, while less massive
%BHs and less luminous late-type hosts are evolved at later
%epoch. Balland et al. (2003) support this scenario: They examined
%merging process of dwarf disk galaxies with a semi-analytical scheme
%to study the evolution of galaxy morphologies along Hubble sequence, and
%suggested that the large spheroid galaxies have been forming stars at
%earlier epoch (their peaks $z \sim 3$) than disk galaxies(their peaks $z
%\sim 1$). 

Since we do not examine host galaxies of AGNs at $z>0.6$, 
it is unknown
%that the hard X-ray luminosity can be indicators of $M_{BH}$s  and 
%properties of host galaxies for higher redshift AGNs. 
whether the BS-relation is hold at the higher redshifts or not.
Studies for estimation of $M_{BH}$s with Kaspi relation 
(Kollmeier et al. 2005) show
that Eddington ratios are between 0.1 -- 1 for AGNs at 
$0.3 < z < 4$, 
%which are comparable to those of AGNs at intermediate redshift, 
and suggest that nuclear
luminosity of AGNs would roughly correlate $M_{BH}$s. 
Meanwhile, BS-relations at higher redshift are rather controversial.
Shields et al. (2003) show that the BS-relation of AGNs does 
not evolve from $z\sim3$ to 0. Akiyama (2005) and Peng et
al. (2005) show that
$M_{BH}$s of higher redshift QSOs ($1 \lesssim z
\lesssim 3$) are larger than those expected from BS-relation in local
universe.
Furthermore, Borys et al. (2005) claim the inverse tendency, i.e.,
$M_{BH}$ is smaller than that expected from BS-relation in local
universe for hard X-ray detected submillimeter galaxies ($1 \lesssim z < 3$).  
Thus, there is no consensus on the BS-relation at the higher redshift.
One of causes of this difference in estimating BS-relation at 
high redshift is a difficulty of decomposition of a spheroid component 
for AGNs.
Making a  sample of type-2 AGNs at such high redshift will 
reduce the difficulty much.
%we could decompose nuclear components of AGNs with smaller
%uncertainty, which would complement studies of BS-relation with type-1
%AGNs: $M_{BH}$s of type-1 objects can be easily estimated by
%Kaspi relation, but bulge luminosity (mass) of these can not be easily
%estimated, and vice versa for type-2 objects. 

In this study, we demonstrate the effectiveness of using type-2
sample in examining hosts, particularly spheroid components,
at intermediate redshifts.
AGN surveys at higher redshifts have been in progress with
$Chandra$ and $XMM-Newton$ satellites and their optical 
and near infrared(NIR) follow-up.
In near future, we will obtain samples of type-2 AGNs in 
high redshift universe.
For such targets, NIR imaging employing (laser guide) adaptive optics 
with ground-based 8-10m class telescopes is expected to be powerful 
tools to study host galaxies of high redshift QSOs.

\acknowledgments

We are grateful to
the staff members of UH 2.2-m telescope. 
 Use of the UH 2.2-m telescope for the observations is supported by
National Astronomical Observatory of Japan (NAOJ).
This research was
supported by a Grant-in-Aid for Scientific Research
from Japan Society for the Promotion of Science (17540216).

\newpage

%\begin{figure}
% \plotone{fig/no26.spec.eps}
% \caption{A spectra of NO26, which is quoted from Akiyama et
% al. (2003). Stellar absorption lines such as $Mg_{b}$, Ca H and Ca K
% are clearly seen in this spectra.\label{spec}}
%\end{figure}

\clearpage
\begin{deluxetable}{cccccccccc}
%{p{2.4cm}p{0.8cm}p{0.8cm}p{1.6cm}p{0.8cm}p{0.8cm}p{0.8cm}p{1.6cm}p{1.6cm}p{0.8cm}}
 \tabletypesize{\scriptsize}
 \tablecaption{Properties of our sample \label{sample}}
 \tablewidth{0pt}
 \tablehead{
 \multicolumn{1}{c}{Name} &
 \multicolumn{1}{c}{ID} &
 \multicolumn{1}{c}{$z$} & 
 \multicolumn{1}{c}{$\log L_{\rm 2-10keV}$\tablenotemark{a}} &
 \multicolumn{1}{c}{$R$} &
 \multicolumn{1}{c}{$M_R$} &
 \multicolumn{1}{c}{FWHM} &  
 \multicolumn{1}{c}{Observed Period} &
 \multicolumn{1}{c}{Instrument} &
 \multicolumn{1}{c}{Photometry} \\
 \multicolumn{1}{c}{} &
 \multicolumn{1}{c}{} &
 \multicolumn{1}{c}{} &
 \multicolumn{1}{c}{[erg s$^{-1}$]} &
 \multicolumn{1}{c}{[mag]} &
 \multicolumn{1}{c}{[mag]} & 
 \multicolumn{1}{c}{[arcsec]} &
 \multicolumn{1}{c}{} &
 \multicolumn{1}{c}{} &
 \multicolumn{1}{c}{}
 }
\startdata
1AXG~J033516$-$1505 & SE20   & 0.122 & 43.28 & 17.46 & -21.39 & 0.78 & Dec/2004 & OPTIC   & (1)  \\
1AXG~J090053$+$3856 & NO54   & 0.229 & 44.02 & 17.88 & -22.36 & 0.84 & Apr/2004 & OPTIC   & (4)  \\
1AXG~J122155$+$7525 & NO40   & 0.239 & 43.69 & 17.40 & -23.12 & 1.0  & Apr/2004 & OPTIC   & (2)  \\
1AXG~J123605$+$2613 & NE04   & 0.459 & 44.37 & 19.50 & -22.89 & 0.93 & May/2005 & Tek2048 & (3)  \\
1AXG~J142353$+$2247 & NE09   & 0.282 & 44.34 & 16.80 & -24.16 & 0.81 & May/2005 & Tek2048 & (2)  \\
1AXG~J144109$+$3520 & NO32   & 0.077 & 42.65 & 16.31 & -21.25 & 0.82 & May/2005 & Tek2048 & (4)  \\ 
1AXG~J144301$+$5208 & NO26   & 0.212 & 43.69 & 17.38 & -22.72 & 1.5  & May/2005 & OPTIC   & (4)  \\
1AXG~J150339$+$1016 & NO05   & 0.095 & 43.29 & 16.35 & -21.80 & 0.77 & May/2005 & Tek2048 & (4)  \\
1AXG~J151524$+$3639 & NO21   & 0.324 & 44.03 & 18.14 & -23.09 & 0.78 & May/2005 & Tek2048 & (4)  \\
1AXG~J160118$+$0844 & NO53   & 0.606 & 44.77 & 20.57 & -23.32 & 0.97 & Apr/2004 & OPTIC   & (4)  \\
1AXG~J163538$+$3809 & NO47   & 0.099 & 43.03 & 17.66 & -20.60 & 1.1  & Apr/2004 & OPTIC   & (3)  \\
1AXG~J170730$+$2353 & NO24   & 0.245 & 44.00 & 19.46 & -21.10 & 0.81 & May/2005 & Tek2048 & (4)  \\
1AXG~J174943$+$6823 & NO43   & 0.051 & 42.81 & 14.96 & -21.85 & 0.72 & May/2005 & Tek2048 & (1)  \\
AX~J130840$+$2955   & LSS014 & 0.164 & 43.21 & 17.48 & -21.98 & 0.95 & May/2005 & Tek2048 & (4)  \\
AX~J131758$+$3257   & LSS195 & 0.214 & 43.40 & 17.58 & -22.59 & 0.90 & May/2005 & Tek2048 & (4)  \\
\enddata

 \tablenotetext{a}{Rest-frame absorption corrected 2--10 keV
 luminosity. Absorption correction was done with the best fit absorbed
 power law model, details are described in Akiyama et al. (2003)}
 \tablerefs{Photometric data derived from (1) this work, (2) Akiyama et al. (2003),
 (3)Watanabe et al. (2004), and (4)SDSS magnitudes with conversion of
 equation (1) in the text.}
\end{deluxetable}

\begin{deluxetable}{cccccc}
 \tabletypesize{\scriptsize}
 \tablecaption{Results of two-dimensional fitting \label{fit_lst}}
 \tablewidth{0pt}
 \tablehead{
 \multicolumn{1}{c}{ID} &
% \multicolumn{1}{c}{$M_R$ (total)} &
 \multicolumn{1}{c}{$M_{\rm sph}$} &
 \multicolumn{1}{c}{$r_{e}$} & 
 \multicolumn{1}{c}{$M_{\rm disk}$} &
 \multicolumn{1}{c}{$r_{d}$} &
 \multicolumn{1}{c}{$M_{\rm PSF}$} \\
 \multicolumn{1}{c}{}&
 \multicolumn{1}{c}{[mag]} &
 \multicolumn{1}{c}{[kpc]} & 
 \multicolumn{1}{c}{[mag]} & 
 \multicolumn{1}{c}{[kpc]} &
 \multicolumn{1}{c}{[mag]}
 }
 \startdata
 SE20   & -20.18 & 3.55  & -20.95 & 5.33   & -18.31 \\
 NO54   & -22.04 & 5.05  & -20.71 & 24.24  & -18.71 \\
 NO40   & -22.49 & 7.26  & -22.11 & 15.24  & -19.55 \\
 NE04   & -22.75 & 5.01  & -19.26 & 23.12  & -20.20 \\
 NE09   & -23.71 & 1.45  & -22.84 & 7.42   & -20.86 \\
 NO32   & -19.80 & 1.09  & -20.90 & 2.46   & -15.33 \\
 NO26   & -22.57 & 10.70 & -19.58 & 20.75  & -19.80 \\
 NO05   & -21.19 & 1.96  & -20.87 & 9.69   & -6.59  \\
 NO21   & -21.88 & 0.84  & -22.60 & 4.46   & -19.22 \\
 NO53   & -22.70 & 2.48  & -22.19 & 9.74   & -20.60 \\
 NO47   & -20.41 & 2.52  & -18.56 & 9.43   & -14.54 \\
 NO24   & -20.24 & 1.81  & -20.24 & 35.79  & -18.49 \\
 NO43   & -17.55 & 0.31  & -21.97 & 2.78   & -17.72 \\
 LSS014 & -21.25 & 4.14  & -20.97 & 4.14   & -19.37 \\
 LSS195 & -22.11 & 3.03  & -21.25 & 7.27   & -19.57 \\
 \enddata
% \tablerefs{}
\end{deluxetable}

\end{document}